\documentstyle[12pt,twoside,fleqn,espcrc1,epsfig]{article}
\newcommand{\s}{$\bar{s}s$}
\newcommand{\p}{$\bar{p}p$}
\newcommand{\f}{$\phi$}
\newcommand{\an}{annihilation}
\newcommand{\ap}{antiproton}
\newcommand{\na}{na\"{\i}ve}
\newcommand{\om}{$\omega$}
\newcommand{\ten}{$f'_2 (1525) $}

\newcommand{\AmS}{{\protect\the\textfont2
  A\kern-.1667em\lower.5ex\hbox{M}\kern-.125emS}}

\title{Experimental Results on Strangeness Production}

\author{M.G.Sapozhnikov\address{Laboratory of Particle Physics,
Joint Institute for Nuclear Research,\\
P.O. Box 141980, Dubna , Russia}}%

\begin{document}
\maketitle

\begin{abstract}
New experimental results on the production of $\phi$ and $f_2'(1525)$
mesons in the annihilation of stopped antiprotons are discussed.
The explanation of these facts in the framework of
the polarized strangeness model is considered.
\end{abstract}

\section{INTRODUCTION}

        The interest to the strange particles production is motivated
by some unexpected results obtained recently on the
role of the strange quarks in the
nucleon.
Intuitively, it is expected
that the \s~ pairs in the nucleon are
not significant, being a component of the nucleon sea quarks.
Indeed, recent analysis of the
parton distributions \cite{Mar.98} shows that the fraction of the
total momentum of the proton carried by strange quarks is 4.6\% at
$Q^2=20~GeV^2$. However, the evaluation of
the $\pi$-nucleon $\sigma$ term shows that
the contribution of the \s~ quarks
to the nucleon mass is by no means negligible,
about 130 MeV \cite{Sai.97}. The analysis of the
lepton deep-inelastic scattering (DIS) data by
the EMC and successor experiments
has indicated that the \s~ pairs in the nucleon are polarized
(for review see \cite{Ell.95b}).
Finally, the strong apparent violation of the
Okubo-Zweig-Iizuka (OZI) rule was
seen in LEAR experiments with stopped antiprotons
(for review see \cite{Sap.95}). The \f~ production
in some reactions of \p~ \an~ at rest exceeds the OZI rule prediction by a
 factor
30-50.

    To explain these unusual experimental results a model was proposed
\cite{Ell.95} based on a nucleon wave function containing
negatively polarized
$s \bar s$ pairs, as suggested by the DIS
experiments. The main aim of this review is to clarify to what extent
the new experimental facts agree with the predictions of this
model.

    It is worthwhile from the very beginning to define which
nucleon strangeness we are interested in. Following Brodsky \cite{Bro.80},
it is useful to distinguish between the extrinsic and intrinsic
nucleon strangeness. Extrinsic \s ~ quarks are generated from
the processes of the QCD hard bremsstrahlung and gluon splitting
$g \to \bar{s}s $. The lifetime of this component is short.
The extrinsic sea quarks obey the QCD evolution equations.
From this point of view the extrinsic strangeness is something which is
under control.

The notion of the intrinsic nucleon strangeness is less self-evident.
It is assumed that the strange quarks created in the nucleon
form a longlived configuration and the
proton wave function can be decomposed as follows:

\begin{equation}
|p>  = a \sum^{\infty}_{X=0}|uud X>   + b \sum^{\infty}_{X=0}|uud \bar{s}s X>
\end{equation}
where $X$ stands for any number of gluons and light $\bar{q}q$ pairs and
the condition $|a|^2 + |b|^2 = 1$ holds neglecting the admixture of
more than one \s~ pair.

The intrinsic nucleon strangeness, in contrast with the
extrinsic strangeness,
should be essentially non-perturbative phenomenon. The question about the
existence of the intrinsic nucleon strangeness is still open. However,
the problems with $\pi N$ $\sigma$--term, possible polarization of
the nucleon strange sea observed in DIS and apparent violation of the
OZI rule
in \ap~ \an~ at rest
 could be regarded as indications on the existence of the
intrinsic nucleon strangeness.

        A reason to expect a non-negligible role of strange quarks
in the nucleon follows from the properties of the QCD vacuum.
From the QCD sum rules calculations \cite{Iof.81},
it is well known that the condensate of the strange
quarks in the vacuum is not small and is comparable with the condensate
of the light quarks:
\begin{equation}
< 0\left|\bar{s}s \right|0> = (0.8\pm0.1)< 0\left|\bar{q}q \right|0>,~~
q=(u,d)
\end{equation}

Thus, the density of $\bar{s}s$ pairs in the QCD vacuum is quite high and
one may expect that the effects of strange quarks in the nucleon will
be also non-negligible.

\section{INTRINSIC NUCLEON STRANGENESS}

        There are different possibilities for the quantum
numbers of \s~ pair in the nucleon.
Let us consider the proton consisting from the
$uud$ and \s~ clusters and assume that the quantum numbers of the
$uud$ cluster is the same as for the proton $J^P = 1/2^+$. Then
some possible
quantum numbers of the \s~ quarks are shown in the
Table~\ref{tab:jpc}.

\begin{table}[hbt]
\newlength{\digitwidth} \settowidth{\digitwidth}{\rm 0}
\catcode`?=\active \def?{\kern\digitwidth}
\caption{The possible
quantum numbers of the \s~ quarks in the nucleon. $\vec{S}$ and $\vec{L}$
are the
total spin and orbital angular momentum of the \s~pair.
$\vec{J}=\vec{L}+\vec{S}$.
The relative angular momentum between the \s~ and $uud$ clusters is $\vec{j}$.}
\label{tab:jpc}
\begin{tabular*}{\textwidth}{@{}l@{\extracolsep{\fill}}rrrrr}
\hline
S & L & j & $J^{PC}$ & State\\
\hline
0 & 0 & 1 & $0^{-+}$ & "$\eta$"\\
1 & 0 & 1 & $1^{--}$ & "$\phi$"\\
1 & 1 & 0 & $0^{++}$ &$^3P_0$ \\
1 & 1 & 0 & $1^{++}$ &$^3P_1$ \\
0 & 1 & 0 & $1^{+-}$ &$^1P_1$ \\
\hline
\end{tabular*}
\end{table}

        One could see that the \s~ could be stored in the nucleon with the
quantum numbers of $\eta$ and $\phi$
if the relative angular momentum
between the \s~ and the $uud$ clusters is $j=1$. But if
$j=0$, then the quantum numbers of
\s~ pair may be
different , including the vacuum quantum
numbers $J^{PC}=0^{++}$.
Predictions of the model will depend drastically
on the assumption about the \s~ quantum numbers.

        It was established earlier\cite{Dov.90} that the existing experimental
data on the production of $\eta$ and $\eta'$ mesons exclude the $0^{-+}$
quantum numbers for the \s~ admixture in the nucleon.

        The assumption that the \s~ pair has
quantum numbers of \f~ also leads to some problems.
 In this case one might expect some additional
\f~ production due to this strangeness, stored in the nucleon. This
quasi-\f~ pair could be easily shaken-out from the nucleon.
Then one should observe the strong apparent violation  of the OZI rule \cite{OZI}.

The OZI rule
predicts that diagrams with disconnected quark lines
should be suppressed compared to connected quark diagrams.
The \s~ strangeonia  (like \f~ or
$f'(1525)$ mesons) should be produced only via
light quark admixture in their wave functions. It means that
the production of strangeonia  is possible only due to the
departure from the ideal mixing. The OZI rule in formulation
of Okubo \cite{Oku.77} strictly forbids formation of \s~ meson
in $pp$ or \p~interaction.
In spite of some deviation from the OZI rule predictions
observed earlier (for
review, see \cite{Sap.95,Ell.95} ), the violation of
the OZI rule does not exceed the 10\% level.

        The situation has changed since a wealth of new high-statistics
data in various $p \bar p$
annihilation channels became available from the experiments at
LEAR.
They provide information on several
final states including $\phi \gamma$, $\phi \pi$, $\phi \eta$,
$\phi \pi \pi$,
$f' \pi$ and $\phi \phi$, in
different experimental conditions which allow
initial-state spin and orbital angular momentum states to be
distinguished.

Anomalously high $\phi$ production was seen in
different channels of annihilation in liquid and gas hydrogen and
deuterium targets
\cite{Rei.91,Ams.95,Abl.94,Abl.95,Abl.95b}.
The highest deviation is for the $\bar{p}p\to \phi\gamma$ channel where
the ratio $R(\phi/\omega)$ between the yields of \f~ and \om~ meson
production is
$R(\phi/\omega) = (243\pm86)\cdot10^{-3}$, i.e. about 50 times larger than the
OZI prediction $R(\phi/\omega)=4.2\cdot10^{-3}$ (for
the quadratic Gell-Mann-Okubo mass formula).

        The most striking feature of the
OZI rule violation found in the experiments at LEAR is its
strong dependence on
the quantum numbers of the initial state.

Thus, the OBELIX collaboration studied the channel
 $\bar{p}p\rightarrow \phi \pi^0$ for \an~ in the hydrogen
targets with different densities \cite{Abl.95b}.
The conservation of $P$ and $C$--parities strictly fixes the possible quantum
numbers of the $\bar{p}p$ initial state to be
either the spin triplet state $^3S_1$, or
the spin singlet  state $^1P_1$.
It was found that for \an~ in liquid, where the $^3S_1$ state is dominant,
the  $\phi\pi^0$ yield is substantial
and the
ratio
   $R(\phi/\omega) = (129\pm35) \cdot10^{-3}~$, by a factor of 30 higher than the
\na~ OZI rule prediction.

At the same time, no
$\phi$'s were found when \an~ took place from
the $ ^1P_1$ initial state.
The same $^3S_1$ dominance had been observed earlier by
the ASTERIX
collaboration \cite{Rei.91}.

    So, not only a large ratio $\phi/\omega$ was found, it turns out
that this ratio miraculously changes depending upon the initial
state. If the \s~ pair
was stored in
the nucleon with the
\f~ quantum number, it is not clear why the shake-out of this pair
depends on the value of the total spin of {\it both} nucleons.

        Moreover, the shake-out of the \s~ pair with \f~ quantum numbers
should lead to the same, "universal" violation of the OZI rule in different
annihilation channels. However, the deviations from the OZI-rule predictions
was found only in some channels. It is not clear
how to explain, for instance, why the annihilation in $\phi \pi$ channel
exceeds the OZI
rule prediction by a factor of 30, but the annihilation in liquid hydrogen
in $\phi \rho$  exhibits no strong violation of the OZI rule.

        To explain these experimental features,
a model of \f~ production was proposed \cite{Ell.95},
based on a nucleon wave function containing negatively polarized
$s \bar s$ pairs.

It was extended to $\Lambda \bar \Lambda$ production
in \cite{Alb.95}, where arguments were given on the basis of
chiral symmetry that the $s \bar s$ pair in the nucleon wave
function might be in a $^3P_0$ state.

\section{POLARIZED INTRINSIC STRANGENESS MODEL}

        Let us consider the production of \s~ strangeonia in
$NN$ interaction
assuming
that the nucleon wave function
contains an admixture of polarized $\bar{s}s$ pairs with $J^{PC}=0^{++}$ and
$1^{++}$ \footnote{I would like to thank V.Markushin for pointing out the
importance of different \s~ configurations.}.

        Then the shake-out of such pairs will not create \f~, but, for
instance, a scalar strangeonium. The concrete candidate on this state is
not firmly established now (see, for discussion \cite{PDG}).
The lowest $\bar{s}s$ scalar seems to be around 1700 MeV.
Therefore the shake-out of the scalar \s~pair from the nucleon
will be
a source of channels with open strangeness, like $\bar{K}K$ and
$KK^*$.

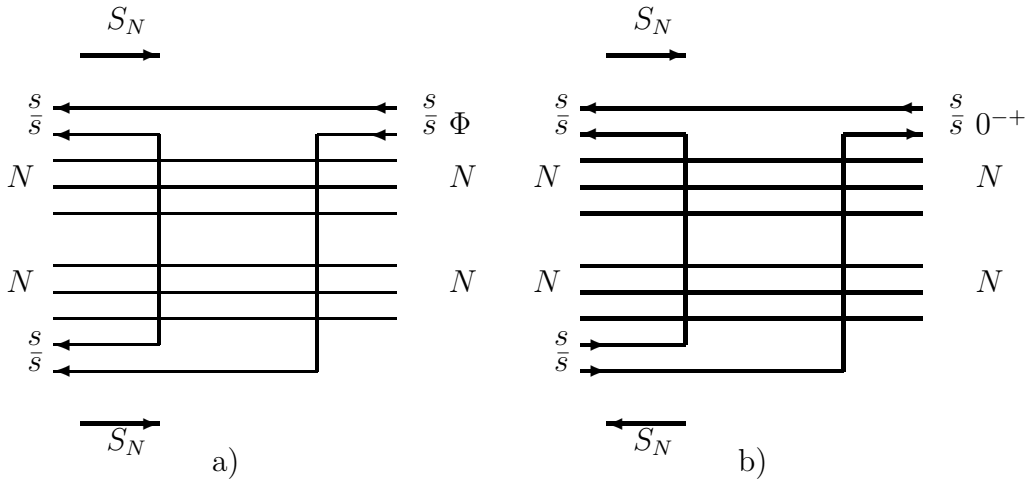
\begin{figure}[htb]
\setlength {\unitlength} {0.7mm} \thicklines
\begin{picture}(180,70)(0,10)
\put(60,20){\vector(-1,0){50}}
\put(30,25){\vector(-1,0){20}}
\put(10,30){\line(1,0){65}}
\put(10,35){\line(1,0){65}}
\put(10,40){\line(1,0){65}}
\put(70,70){\vector(-1,0){60}}
\put(30,65){\vector(-1,0){20}}
\put(10,60){\line(1,0){65}}
\put(10,55){\line(1,0){65}}
\put(10,50){\line(1,0){65}}
\put(30,65){\line(0,-1){40}}
\put(60,20){\line(0,1){45}}
\put(75,70){\vector(-1,0){5}}
\put(75,65){\vector(-1,0){5}}
\put(60,65){\line(1,0){10}}
\put(75,35){\line(-1,0){20}}
\put(75,30){\line(-1,0){20}}
\put(40,1){a)}
\put(5,20){$\bar{s}$}
\put(5,25){$s$}
\put(5,65){$\bar{s}$}
\put(5,70){$s$}
\put(80,65){$\bar{s}$}
\put(80,70){$s$}
\put(85,65){$\Phi$}
\put(1,35){$N$}
\put(1,55){$N$}
\put(85,35){$N$}
\put(85,55){$N$}
\put(20,85){$S_{N}$}
\put(20,5){$S_{N}$}
\linethickness{0.5mm}
\put(15,10){\vector(1,0){15}}
\put(15,80){\vector(1,0){15}}


\put(110,20){\vector(1,0){5}}
\put(160,20){\line(-1,0){45}}

\put(110,25){\vector(1,0){5}}
\put(130,25){\line(-1,0){15}}

\put(110,30){\line(1,0){65}}
\put(110,35){\line(1,0){65}}
\put(110,40){\line(1,0){65}}
\put(170,70){\vector(-1,0){60}}
\put(130,65){\vector(-1,0){20}}
\put(110,60){\line(1,0){65}}
\put(110,55){\line(1,0){65}}
\put(110,50){\line(1,0){65}}
\put(130,65){\line(0,-1){40}}
\put(160,20){\line(0,1){45}}
\put(175,70){\vector(-1,0){5}}
\put(170,65){\vector(1,0){5}}
\put(160,65){\line(1,0){10}}
\put(175,35){\line(-1,0){20}}
\put(175,30){\line(-1,0){20}}
\put(140,1){b)}
\put(105,20){$\bar{s}$}
\put(105,25){$s$}
\put(105,65){$\bar{s}$}
\put(105,70){$s$}
\put(180,65){$\bar{s}$}
\put(180,70){$s$}
\put(185,65){$0^{-+}$}
\put(101,35){$N$}
\put(101,55){$N$}
\put(185,35){$N$}
\put(185,55){$N$}
\put(120,85){$S_{N}$}
\put(120,5){$S_{N}$}
\linethickness{0.5mm}
\put(130,10){\vector(-1,0){15}}
\put(115,80){\vector(1,0){15}}
\end{picture}
\caption{Production of the \s~ mesons in $NN$ interaction
from the spin-triplet (a) and spin-singlet (b) states.
The arrows
show the direction of spins of the nucleons and strange quarks.}
\label{fig:8}
\end{figure}

The \f~ should produced due to a process where
strange quarks from {\it both} nucleons are participating. An example of
such rearrangement diagram is shown in Fig.~\ref{fig:8}.

        If the nucleon spins are parallel (Fig.~\ref{fig:8}a),
then the spins of
the $\bar{s}$ and $s$ quarks in both nucleons are also parallel.
If the polarization of the strange quarks is not changed during the
interaction, then the $\bar{s}$ and $s$ quarks will have parallel
spins in the final state. The total spin of \s~ quarks will be $S=1$ and
if their relative orbital momentum is $L=0$, it means that the
strangeonium has the \f~ quantum numbers, if $L=1$, it will correspond to
the creation of tensor strangeonium, $f'_2(1525)$.

        If the initial $NN$ state is a spin-singlet, the spins of
strange quarks in different nucleons are antiparallel and
the rearrangement diagrams like that in Fig.\ref{fig:8}b may lead
to the \s~ system in the final state with total spin $S=0$. It means
that for $L=0$ a strangeonium with the pseudoscalar quantum numbers
$0^{-+}$ is produced.
One may expect that the formation of $\eta$ meson
will be enhanced from the spin-singlet state.

        The model predicts that the energy dependence of the
$\phi$ production should follow the percentage of the $^3S_1$ state.
It means that in \ap~\an~in flight the \f~ yield will decrease with
increasing of the antiproton energy.

        It is important to note that these rules should hold
as for antiproton-proton annihilation, as for nucleon-nucleon
interaction.

Therefore, the predictions of the polarized strangeness model
are quite definite:
\begin{itemize}
\item the \f~ should produce mainly from the $^3S_1$ state
\item the \ten~  should produce mainly from the $^3P_J$ states
\item the spin-singlet initial states favour the formation of
pseudoscalar strangeonia.
\end{itemize}

It is also quite straightforward to
consider formation of $\bar{\Lambda}{\Lambda}$ and $\phi \phi$
systems.

        Let us confront the predictions of the model
with the new experimental results.

\section{EXPERIMENT AND THE POLARIZED STRANGENESS MODEL}

\begin{figure}[hb]
\psfull
\epsfig{file=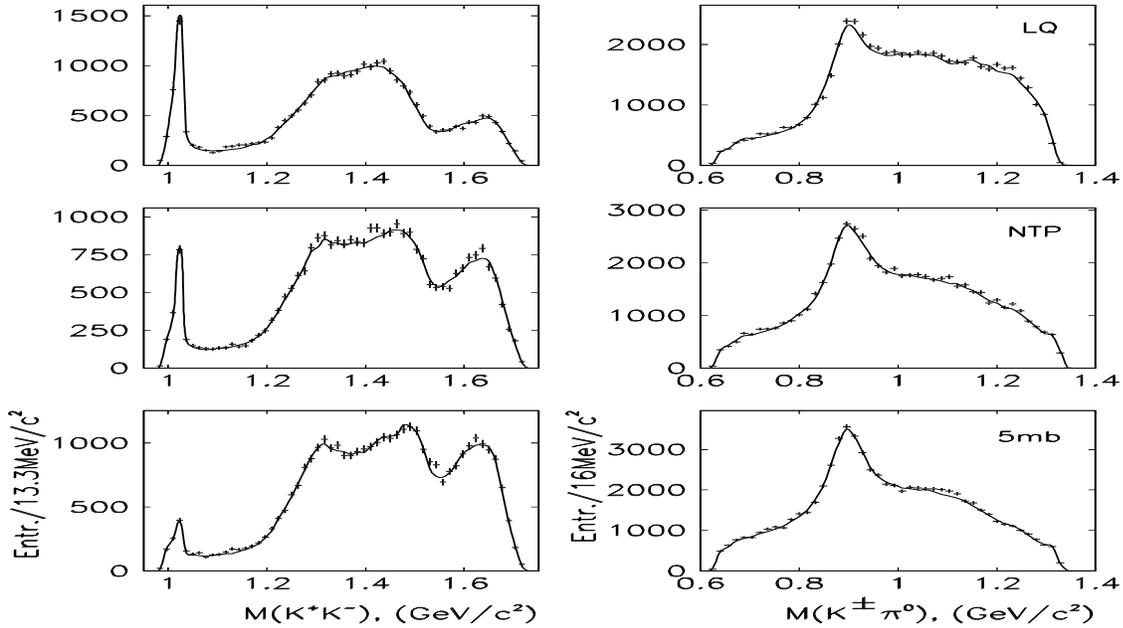,width=15.cm,height=8.cm,
bbllx=0.cm,bblly=0.cm,bburx=16.cm,bbury=17.cm}
\caption{
 $K^{\pm}\pi^0$ and $K^+K^-$ invariant mass distributions for
 $\bar{p}p \to K^+K^-\pi^0$ annihilations in the $H_2$ target:
 liquid (top), gas at NTP (middle), and 5 mbar (bottom).}
 \label{fig:compfit}
\end{figure}

    Recently the measurements of the $\bar{p}p \to K^+K^-\pi^0$
channel for annihilation of stopped antiprotons in liquid,
gas at NTP and
5 mbar pressure, was performed by the OBELIX collaboration
\cite{Pra.98}.
The
invariant mass distributions of the $K^+K^-$ and $K^{\pm}\pi^0$ systems
are
shown in fig.~\ref{fig:compfit}.

        One could see that the peak from the $\phi$ meson
reduces with decreasing of the density of the target whereas
the part of the $K^+K^-$ spectra with high invariant
masses $M > 1.5$~GeV/$c^2$
is more prominent for the low pressure data.

        The dependence of the \f~ yield on the density clearly indicates
the dominance of the production from the $^3S_1$ state.
Using the parameters of the  \p~ cascade
from \cite{Batty}, it is possible to evaluate from the OBELIX data
\cite{Pra.98} the branching ratios of the
$\bar{p}p \to \phi\pi^0$ channel for definite initial states:
\begin{eqnarray}
 Br(\bar{p}p\to \phi\pi^0, ~^3S_1) & = & (7.57\pm0.62)\cdot 10^{-4}~,
\label{3s}\\
 Br(\bar{p}p\to \phi\pi^0, ~^1P_1) & < & 0.5\cdot 10^{-4} ~~~~~~~~~~~~~~
 \mbox{, with~95\%~CL}~ \label{1p}
\end{eqnarray}

Therefore, the indication  of
 strong dependence of the $\phi\pi^0$ production on
 quantum numbers of the initial $\bar{p}p$ state,
obtained by the ASTERIX collaboration
\cite{Rei.91}, is confirmed with
the statistics of factor 100 higher.
 The branching ratio of the $\phi\pi^0$ channel from
 the $^3S_1$ initial state is at least by 15 times larger than that
 from the $^1P_1$ state. There is no theoretical model except
the polarized nucleon strangeness approach to explain this
remarkable selection rule.

        It is interesting to compare characteristics of the
$\bar{p}p \to \phi \pi$ channel with the
$\bar{p}p \to \omega \pi$  ones. Does the same selection rule
exist also for the $\omega \pi$ final state? New results from the
OBELIX collaboration \cite{Don.98,Fil.98}
presented at this Conference do not confirm this guess.

It turns out that for $\omega \pi$ channel the branching ratio
of \an~ from $^1P_1$ state is not negligible. It leads to
different dependences of the $\phi \pi$ and $\omega \pi$
\an~frequencies on the target density. Preliminary results \cite{Don.98}
for the
measurements of the ratio $R = Y(\phi\pi)/Y(\omega \pi)$ are:
$R = (114\pm10)\cdot 10^{-3}$ for \an~ in liquid  and
$R = ( 83\pm10)\cdot 10^{-3}$ for \an~ in  gas at NTP.

        The difference between $\phi \pi$ and $\omega \pi$ channels
is even more pronounced in the measurements of antineutron-proton \an~
\cite{Fil.98}. The cross sections of the
$\bar{n}p \to \phi (\omega) \pi^+$ channels was measured
for antineutron momenta
50-405 MeV/c. It turns out that the
$\phi \pi^+$
cross section drops with energy, strictly following the decreasing of
  S-wave. The
$\omega \pi^+$
cross section has different energy dependence with the contribution
of $^1P_1$ component
  as large as $34\pm7$ \% !

        It seems that the production mechanisms of the
$\bar{N}N \to \phi \pi$  and
$\bar{N}N \to \omega \pi$ reactions  near the threshold are quite
different. As a result the $\phi/\omega$ ratio is not constant with
the energy of antiproton but decreases.

        Probably this effect of  decreasing of the $\phi/\omega$
ratio at high energies
is responsible for the result obtained by the DISTO collaboration
in the measurements of the \f~ and \om~ production in $pp$ interactions
\cite{Bed.98}.
At the proton energy of 2.85 GeV, i.e. at 83 MeV above the \f~
production threshold, it was found that
\begin{equation}
R= \frac{\sigma(p p \to pp\phi)}{\sigma(p p \to pp\omega)} =
   (3.0 \pm 0.5 ^{+1.0}_{-0.8})\cdot10^{-3} \label{pp}
\end{equation}
        One may interpret this fact as a dilution of the
S-wave spin-triplet initial state at high energies. However, direct
experimental measurements of this ratio near the threshold are badly
needed.

        Important information on the dynamics of strangeonia
comes from the OBELIX analysis of the formation of the tensor \s~meson \ten~
\cite{Pra.98}.
    From the OZI--rule it is expected that the ratio between the yield of
\ten~ production and that of $f_2(1270)$ meson, which consists from
the light
quarks only,
is on the level of
$R(f'_2 /f_2)=(3-16)\cdot10^{-3}$.
 The measurements of
the $K^+K^-\pi^0$ channel at three hydrogen densities \cite{Pra.98}
provide a possibility
to determine this ratio for \an~ from the S- and P-wave. It turns out
that
\begin{eqnarray}
 R(f_2'(1525)\pi^0/f_2(1270)\pi^0)
 & = &(47 \pm 14 ) \cdot 10^{-3} ~,~~~~\mbox{S-wave}  \label{f5} \\
 & = &(149 \pm 20 ) \cdot 10^{-3} ~,~~~\mbox{P-wave}  \label{f6}
\end{eqnarray}

        Indeed, the strong apparent violation of the OZI rule is seen
just for
\an~ from the P-wave, as predicted by the polarized strangeness model.

        Another interesting result concerns the measurement of
the $\bar pp$ annihilation at rest into the $\phi\eta$
final state for
liquid hydrogen, gas at NTP and at a low pressure  of $5$ mbar,
which was performed by the OBELIX collaboration \cite{Nom.98}.
The $\phi\eta$ final state has the same $J^{PC}$ as the
$\phi\pi^0$ final state. So, one may expect to see the same
decreasing of the \f~ yield with the target density, as it was
observed  for the $\phi \pi^0$ channel
(see, Fig. \ref{fig:compfit}).
However, unexpectedly, the reverse trend is seen:
the yield of the $\bar{p}p \to \phi \eta$ channel grows
with decreasing of the target density.

Using the same parameters of \p~ atom cascade for the evaluation
 of the branching ratios as in
\cite{Pra.98}, it is obtained that
\begin{eqnarray}
B.R.(\bar{p}p \to\phi\eta, ^3S_1) & = &(0.76\pm 0.31)\cdot10^{-4} \label{r1} \\
B.R.(\bar{p}p\to\phi\eta, ^1P_1) & = &(7.72\pm 1.65)\cdot10^{-4}  \label{r2}
\end{eqnarray}

One should compare these results with those of (\ref{3s}-\ref{1p}).
Again we see a strong dependence of the yield on the initial state quantum
numbers.

        The interpretation of the $\phi \eta$ production in the
framework of the polarized intrinsic strangeness model is not
straightforward.
Since $\eta$ meson has a substantial $\bar{s}s$ component, the
production of the $\phi \eta$ final state could be regarded as
the production of two $\bar{s}s$ pairs, one in the spin triplet state
and the other in the spin singlet state.

        If we treat the reaction $\bar{p}p \to \phi\eta$ as formation
of pseudoscalar \s~ strangeonium, then the polarized
intrinsic strangeness model predicts that it should be formed from the
spin singlet initial state.
It is interesting that the same
strong enhancement of the $\eta$ production from the initial spin singlet
state was observed in
$pp\to pp\eta$ and $pn\to pn\eta$ reactions \cite{Cal.98}. An attempt
to interpret this effect in the polarized nucleon strangeness model
was done in \cite{Rek.97}. They pointed out that at threshold
the ratio between $\eta$ production on neutron
and on proton is quite simple:
\begin{equation}
R_{\eta}=\frac{\sigma(np \to np\eta)}{\sigma(pp \to pp\eta)} =
\frac{1}{4} (1+ \frac{|f_0|^2}{|f_1|^2})             \label{reta}
\end{equation}
where $f_1$ and $f_0$  are the amplitudes corresponding to the
total isospin $I=1$ and $I=0$, respectively. At threshold,
when the orbital momentum of
two nucleons in the final state is $l_1=0$ and the orbital momentum of the
produced meson relative to the center of mass system of these two
nucleons is also $l_2=0$, the connection between the isospin and the total
spin of two nucleons in the initial state is fixed.
The amplitude $f_1$ corresponds to the spin-triplet
initial nucleon state  and
the amplitude $f_0$ corresponds to the spin-singlet one.
Therefore, using the experimental data on the $pp$ and $np$ cross sections,
it is possible to estimate
the ratio between spin-singlet and spin-triplet amplitudes.

	Recent measurements of the
$\eta$ production in the threshold region \cite{Cal.98} show that
the ratio is fairly constant at approximately $R_{\eta} \approx 6.5$.
It means from (\ref{reta}) that the spin-singlet amplitude dominates
$ |f_0|^2/|f_1|^2 \approx 25$.

It resembles the dominance of the $\eta$ formation from
the spin-singlet initial state observed in \p~ \an, where from
(\ref{r1})-(\ref{r2}) it follows that

\begin{equation}
 \frac{Y(\bar{p}p\to\phi\eta;S=0)}{Y(\bar{p}p\to\phi\eta;S=1)} = 9.2\pm2.0~        \label{ret2}
\end{equation}

        The polarized strangeness model explains not only the \s~ meson
production. In \cite{Alb.95} it was extended to
$\bar{p}p \to \Lambda \bar \Lambda$ channel. The PS 185 experiment
\cite{Tor.98}
observe a remarkable absence of spin-singlet fraction $F_s$
in the $\Lambda \bar \Lambda$ final state
$ F_s = 0.00014\pm0.00735$.
That is in agreement with the polarized strangeness model expectations.
If the strange \s~ quarks were polarized in the initial state,
it is natural to expect that they will keep the total spin $S=1$ in the
final state.

The PS 185 collaboration has also measured  the
$\bar{p}p \to \Lambda \bar \Lambda$ channel
 for annihilation on the polarized target to
evaluate the target spin depolarization $D_{nn}$. The polarized strangeness
model predicts \cite{Alb.95} that the $D_{nn}$ value is negative.

        The JETSET collaboration has seen unusually high apparent violation
of the OZI-rule in the
 $ \bar{p} + p \to \phi + \phi$
channel \cite{Lov.97}, \cite{Eva.98}. The measured cross section of
this reaction
turns out to be 2-4 $\mu$b  for the momenta of incoming antiprotons from
1.1 to 2.0 GeV/c. It is by two orders of magnitude higher than the value
of 10 nb expected from the OZI-rule.
If this apparent OZI violation is due to presence of the polarized
strangeness in the nucleon, then
it was predicted \cite{Ell.95}
that the $\phi \phi$ system
should be produced mainly from the initial
spin--triplet state.
Indeed, recent data of the JETSET collaboration
\cite{Lov.97} have demonstrated that the initial spin--triplet state with $2^{++}$
is dominated.

Moreover, preliminary analysis \cite{Lov.95} shows that
the final states with the total spin $S$ of \f\f~ system  $S=2$ are
enhanced. This fact could be naturally explained in the polarized
strangeness model with the
same arguments  as spin--triplet dominance of $\Lambda \bar{\Lambda}$
system created in \p~ \an~ \cite{Alb.95}.

        Of course, the polarized nucleon strangeness model is not
the only possible explanation of the facts. In case of the \f\f~ production
the "simplest" one is that the $2^{++}$ dominance is the signal of a
tensor glueball. The absence of the spin--singlet state in the
$\Lambda \bar{\Lambda}$ system could be reproduced in meson-exchange
models (see, for instance, \cite{Hai.92}).
 The anomalously high yield of the $\bar{p}p \to \phi\pi^0$
channel could be explained (see \cite{Loc.94,Lev.94,Gor.96})
 by the rescattering diagrams with the OZI-allowed transitions
in the intermediate
state, for instance,  $\bar{p}p \to K^*\bar{K} \to \phi\pi^0$.
 The calculations \cite{Loc.94,Lev.94,Gor.96} provide a reasonable agreement
(within a factor of two) with the experimental data on
 the $\phi\pi$ yield for
annihilation from the S-wave.

        However, what is not explained today is the reason of the
strong dependence of the \f~ yield on the spin of the initial state.
In the conventional approaches, without assumption about the polarized
nucleon strangeness, it is unclear  why the
$\phi\pi$ yield for annihilation from the S-wave is so strong but
from the P-wave it is absent. Even more strange that for the $\phi\eta$ channel
the situation is reversed: P-wave is dominant and the \an~ from the
S-wave is suppressed.

Unexplainable in the standard mechanisms
is the opulent production of the tensor strangeonia from the P-wave.
The production of $f'_2$ in the
$\bar{p}p \to f'_2 \pi^0$ reaction
was calculated in \cite{Lev.95} via final state interactions of
$K^*K$ and $\rho\pi$.
The obtained production rates of $f'_2$
are rather small, about $10^{-6}$. That is by two orders of magnitude less
than the values measured by the OBELIX collaboration \cite{Pra.98}.

\section{FUTURE EXPERIMENTS}

       A number of interesting experiments could be proposed searching for the
effects of nucleon intrinsic strangeness. For instance, it is important
to verify if the selection rules found in
the \ap~ \an~ at rest exist also for the nucleon-nucleon or electron-nucleon
interactions. Thus, at the ANKE spectrometer at COSY it is planned
\cite{COSY} to measure
the \f~ production in the polarized proton interactions with the polarized proton
target
\begin{equation}
\vec{p} + \vec{p} \to p + p + \phi
\end{equation}
If the \f~production
in the nucleon-nucleon interaction is dominated by the spin--triplet
amplitude, as was observed in  the \ap~ \an~, then the \f~production should be
maximal when the beam and target nucleons have parallel polarization and
suppressed when they are antiparallel.

	It is possible also to verify the spin dependence of the \f~
production amplitude using non-polarized nucleons.
The \f~ production in $np$ and $pp$ collisions
at threshold should also follow Eq.(\ref{reta}).
If \f~ is not produced from the spin--singlet states,
then the ratio of the $np$ and
$pp$ cross sections at threshold is
\begin{equation}
R_{\phi}=\frac{\sigma(np \to np\phi)}{\sigma(pp \to pp\phi)} =
\frac{1}{4} (1+ \frac{|f_0|^2}{|f_1|^2}) \approx \frac{1}{4}  \label{rfi}
\end{equation}
Remarkably that recently
this ratio was calculated \cite{Tit.97} in the framework of
one-boson exchange model, i.e. without assumption about the
nucleon intrinsic strangeness.  It turns out to be $R_{\phi}=5$.
Therefore experimental measurements of this ratio near the threshold
could discriminate between the predictions of these theoretical models.

	An interesting programme of measurements which could verify the
nucleon intrinsic strangeness is planned by the
COMPASS collaboration \cite{COMPASS} at CERN. 
The measurements of
the deep inelastic scattering
of polarized muons on polarized target are planned to study contributions
from quarks and gluons to the nucleon spin. The intrinsic strangeness model
predicts \cite{Ek.95} that the $\Lambda$ hyperons created in the target
fragmentation region should have large negative longitudinal polarization.
It will be possible to verify this prediction in the COMPASS experiment
with a large statistics of the order of $10^5$ $\Lambda$ hyperons.

       	As we see, up to now there are no experimental facts which
could be ruled out the intrinsic nucleon strangeness hypothesis.
It gives credence to the approach and stimulates further investigations.

\section{Acknowledgements}

	Deep thanks for fruitful collaboration to all friends and
colleagues in the OBELIX experiment
at LEAR.
 	Numerous discussions with
J.Ellis, A.Kotzinian, D.Kharzeev, F.Lev and A.Titov
are highly appreciated.  Special thanks to V.Markushin and M.Locher for
the extremely important comments to the draft of the contribution.
I would like to thank the Organizers for warm hospitality at
Villasimius and the financial support.

\end{document}